\documentclass[prb,preprint,floatfix]{revtex4-1} 

\usepackage{amsmath} 
\usepackage{amsfonts} 
\usepackage{graphicx} 

\usepackage{soul}
\usepackage{color}

\begin{document}


\title{Stationarity and energy transfer in out-of-equilibrium systems }
\author{S\'ebastien Auma\^\i tre$^{1,2}$ Antoine Naert$^1$}
\email[Corresponding author. Email address: ]{sebastien.aumaitre@cea.fr}
\affiliation{$^1$Laboratoire de Physique, ENS de Lyon, UMR-CNRS 5672, 46 all\'ee d'Italie, F69007 Lyon, France \\
$^2$Service de Physique de l'Etat Condens\'e, DSM, CEA-Saclay, UMR 3680 CEA-CNRS, 91191 Gif-sur-Yvette, France}

\begin{abstract}

This work proposes a definition for a characteristic energy density based on the measurement of the two first moments of the extrinsic injected power smoothed over time. Using the properties of stationary processes, we show that this definition also characterizes an energy per degrees freedom of the intrinsic dissipative processes. In contrast to previous approaches, our framework does not necessitate explicitly a contact with a thermostat. Hence, it can be applied as usual to systems in contact with thermostats put out of equilibrium by an external driving. Yet, our approach holds also for intrinsically dissipative macroscopic systems that go at rest when the forcing is stopped. Here we focus on the average and the variance of the fluctuations. We are not concerned about the fluctuations around zero of the smoothed injected power that can be extremely rare and difficult to catch experimentally. Then we show that the characteristic energy density we defined, reduces to the kinetic energy of a Brownian-like particle described by a set of Langevin equations with a viscous damping term. The {\it particle} can be either in contact with a thermostat or intrinsically dissipative and driven by a random force. In the first case, our formalism allows us to recover the result obtained in the framework of the fluctuation relation but it extends it for a correlated thermal noise. Our characteristic energy density is measured in an experimental system where nonlinear waves are generated in a thin elastic plate by a large shaker. A smaller shaker attached to the moving plate is used as a probe to measure the energy exchanged with the plate excited by the large shaker. For both shakers, the proportionality of our characteristic energy density with the kinetic energy is demonstrated. It is a consequence of the viscous damping driving the dissipation in this system. Another system with nonlinear frictional dissipation is also investigated. We integrate a numerical model consisting of a set of blocks attached together by springs and the first being driven at constant speed. We show that in this case, our definition of energy density deduced from fluctuations of injected power still characterizes the dissipation but is not more proportional to the kinetic energy because the dissipative process is not a viscous damping anymore. 

\end{abstract}

\maketitle
\section{Introduction}
\label{intro}
{One can consider two distinct classes of dissipative systems. The first includes systems in contact with a thermostat, forced out of equilibrium by an external stimulus. At zero forcing these are in equilibrium with their thermostat. The second class gathers intrinsically dissipative {\it athermal} systems. It includes all the macroscopic systems in which the {\it thermal degrees of freedom} are not resolved. The system goes at rest when the forcing is stopped.
A theoretical framework have been developed during the $90$s to study the first class. The entropy creation rate is deduced from the measurement of the work applied during a time lag $\tau$, asymptotically long compared to any microscopic time of the system. The ratio of the probability to observe a fluctuation of entropy creation rate $\sigma_\tau$ during a time $\tau$ over the probability to observe the same fluctuation with the opposite sign, is related to the exponential of $\sigma_\tau$ times $\tau$ divided by the Boltzmann constant $k_B$, i.e.:
\begin{equation}
\frac{P[\sigma_\tau]}{P[-\sigma_\tau]}=\exp\left(\frac{\tau\sigma_\tau}{k_B}\right)
\label{FluctautionTh}
\end{equation}
This is the so call {\it Fluctuation Relation} (FR) that inspired several further developments \cite{EvansCohenMoriss,GallavottiCohen,Jarzynski,Crooks,Sasa}. It has have been first evidenced on a sheared gas simulated by molecular dynamics \cite{EvansCohenMoriss}. In the simulation of such a forced system, one has to mimic the action of thermostat to prevent the system warm up. It is done via an adaptive damping insuring an instantaneous balance between injected and dissipated power which lets the microscopic dynamics reversible. In that sense, it relates the extrinsic applied power smoothed over a time $\tau$, $I_\tau$, to the intrinsic entropy creation rate which is directly connected to the dissipation by $\sigma_ \tau=D_\tau/T$ in that context, with $T$ the temperature of the thermostat. This fluctuation relation became a {\it theorem} for a special class of chaotic systems where the microscopic reversibility is also preserved and the instantaneous balance of powers is prescribed \cite{GallavottiCohen}. Then the {\it Fluctuation Theorem} has been successfully demonstrated for Brownian particles \cite{SearlesEvans,KurchanLect}. There is no such theoretical framework for the second class of dissipative systems. Nevertheless, in the specific case of intrinsically irreversible system described by a Langevin-like equation, exact computations of the injected and dissipated powers fluctuations are possible \cite{Farago1}. We propose in this work an analytic tool to characterize some properties of the intrinsic dissipation processes from the measurement of the extrinsic injected power in both classes of dissipative systems. Our tool is based only on fundamental properties of stationary processes and expresses as the FR in the Gaussian limit.

Experimentally the Fluctuation Relation has been tested on many kinds of non-equilibrium systems: either on small size systems in contact with equilibrium thermostats \cite{Bustamante2,Bustamante1,Ciliberto1,Garnier,SergioArtem}, or devices in contact with an out-of-equilibrium energy reservoir \cite{SergioArtem,Naert,Bandi}. Moreover, it has been tested on intrinsically dissipative systems \cite{Laroche,Kellay,Saumaitr}. It must be noticed that these tests are quite difficult. Indeed the Fluctuations Relation (\ref{FluctautionTh}) deals with fluctuations around 0 of the work applied during a time $\tau$ or equivalently to the injected power smoothed over this time $\tau$. Rigorously the FR holds only in the very long time limit. Because the theory of large deviation, the fluctuations around zero of such smoothed variable become at least exponentially rare. Hence, the experimental tests require assumptions on the shape of the probability density function \cite{Laroche,Bustamante2,Kellay}, or a fine adjustment of system parameters in order to get an injected power with very a small average value and very large fluctuations to insure enough negative fluctuations\cite{Garnier,Naert}. In many situations, the range where relation (\ref{FluctautionTh}) can be checked is very limited due to the lack of statistical convergence. This often prohibits definitive conclusions about the relevance of the fluctuation relation in realistic systems \cite{Saumaitr}. The approach we propose in this article avoids this pitfall. Indeed we are interested in typical fluctuations which are more relevant than rare events around zero. Only the two first moments of the smoothed injected power are necessary. We establish a moment convergences formula relating these measurements to characteristic fluctuations of the dissipation. It allows us to define a characteristic energy density, intensive with respect of the relevant degrees of freedom pertinent for the dissipation.\\

In the following, we first recall the strong consequences of the stationarity on the fluctuations of the injected and dissipated powers in \ref{Stat}. Then, in section \ref{Fluct}, we gives the expression of the Fluctuation Relation (\ref{FluctautionTh}) expressed with the injected power assuming Gaussian fluctuations. It involves only the two first moments. Putting all this together, one can relate the mean, the variance and the correlation time of the dissipated power. This is what we call the relation of the moments convergence, which also defines a characteristic energy density. The subsection \ref{ThCL} shows how this relation can be interpreted in the framework of the central limit theorem. In the section \ref{Langevin}, we apply this expression to a Brownian particle described by the Langevin equation. Two cases are considered. In the first one, \ref{C1}, the particle, in contact with a thermostat, is pushed out of its equilibrium. Hence, it belongs to the first class of dissipative system, as defined above. Results obtained by previous authors for the Langevin equation is recovered\cite{KurchanLect}. In addition, we propose a correction to account for a finite time correlation on the forcing. 
In \ref{C2}, we consider an athermal system, in the sense that a particle intrinsically dissipative is randomly excited by an external forcing in the absence of thermal bath. It therefore belongs to the second class.
In both cases, a characteristic energy density can be deduced and related to the particle kinetic energy via the viscous damping. Section \ref{Exp} is completely devoted to experiments. First, in \ref{ExpDevice}, the setup is presented. It is constituted of a thin elastic plate forced in a turbulent state by a large shaker. A second shaker, smaller, is used as a probe in contact with the plate. This excited plate is considered simply as an out-of-equilibrium reservoir of kinetic energy. Section \ref{Result} is devoted to the experimental results. We show that the relation of the moments convergence of exhibits a characteristic energy density proportional to the variance of the speed of the shakers. It is expected in our formalism, which assumes a viscous damping. However, we show that the small shaker is biased by the power dissipated in its internal resistance. Finally, in section \ref{BK}, we apply the relation of the moments convergence to a system with solid friction damping. We simulate a Burridge-Knopoff set of three blocks linked together with springs, sliding with nonlinear friction, the first block being pulled at constant velocity. In that case, the relation of the moments convergence still holds. However, the characteristic energy density defined by this relation cannot be related to the velocity variance of the blocks. Indeed, in our approach, such a connection is expected only for viscous damping. Some conclusions and perspectives are drawn in the last section \ref{Conclu}.

\section{The convergence of the moments in out-of-equilibrium systems}
\label{moment}
\subsection{Stationarity in out-of-equilibrium system}
\label{Stat}

Let us consider an out-of-equilibrium system such that:
\begin{equation}
\frac{dE(t)}{dt}=I(t)-D(t)
\label{EnBal}
\end{equation}
with $E$ the internal energy, $I$ the injected power and $D$ the dissipated power. The stationarity imposes that, in average, $\langle I \rangle=\langle D \rangle$ where $\langle~.~ \rangle$ denotes an ensemble average. But it also implies \cite{Farago}:
\begin{equation}
\int_0^{+\infty} \langle I(t)\; I(t+\tau) \rangle d\tau=\int_0^{+\infty} \langle D(t)\; D(t+\tau) \rangle d\tau
\label{Intxcorr}
\end{equation} 
which does not depend on $t$t for stationary processes. It can be rewritten as:
\begin{equation}
t_I\;\sigma(I)^2=t_D\; \sigma(D)^2
\label{IntxcorrII}
\end{equation}
with the usual definition of the correlation time of $X$: 
$t_X=\frac{1} {\sigma(X)^2}\; \int_0^{+\infty} \langle X(t)\; X(t+\tau) \rangle d\tau $, where $\sigma(X)^2$ is the variance of $X$. The relations (\ref{Intxcorr}) and (\ref{IntxcorrII}) express the convergence of the fluctuations of the smoothed variable $I_\tau$ and $D_\tau$ at large $\tau$. Here the smoothed variables are defined as 
$X_\tau(t)=\frac{1}{\tau}\;\int_t^{t+\tau} X(t')dt'$. Due to the Wiener-Kinchine theorem, equation (\ref{Intxcorr}) implies also the convergence of the Power Density Spectrum of $D$ and $I$ at vanishing frequency. As far as low frequencies are concerned, the balance between injected and dissipated power is recovered. Actually, after suitable integration one has \cite{Farago}:
\begin{eqnarray}
\sigma(I_\tau)^2=&\frac{1}{\tau}\int_0^{+\infty} \langle I(t)\; I(t+\tau) \rangle d\tau&=\frac{t_I}{\tau}\;\sigma(I)^2\\
=&\frac{1}{\tau}\int_0^{+\infty} \langle D(t)\; D(t+\tau) \rangle d\tau&=\frac{t_D}{\tau}\;\sigma(D)^2\\
=&\sigma(D_\tau)^2&
\label{sigmaItau}
\end{eqnarray}
for any large $\tau \gg t_I$ and $\tau \gg t_D$ at the leading order in $1/\tau$.
An instantaneous balance of power was introduced in the first highlighting of the Fluctuation Relation and it is a requirement in the first demonstrations of the Fluctuation Theorem \cite{EvansCohenMoriss,GallavottiCohen}. Here we do not require an instantaneaous balance but we show that it occurs surely for the lowest frequencies since Power Density Spectrum of $D$ and $I$ converge together at vanishing frequency.

\subsection{Application to the Fluctuation Relation with Gaussian statistics} 
\label{Fluct}
The Fluctuation Theorem relates the asymmetric function:
\begin{equation}
\label{AssFundef}
f(\epsilon)\doteq\frac{1}{\tau}\log\frac{P(I_{\tau}=\epsilon)}{P(I_{\tau}=-\epsilon)}
\end{equation}
to the temperature of the thermostat surrounding the system via:
\begin{equation}
\label{ThFlAsFun}
f(\epsilon)=\frac{\epsilon}{k_{\rm B}T}
\end{equation}
with $k_B$ the Boltzmann constant. \\
One can rewrite the preceding equality with the {\it characteristic energy density} $E_c$ : $f(\epsilon)=\frac{\epsilon}{ E_c}$. 
many authors \cite{Kellay,Naert,Eckhardt04} underline that the definition (\ref{AssFundef}) imposes for Gaussian fluctuations of $I_\tau$: $E_c\dot=\frac{\tau}{2} \frac{\sigma(I_\tau)^2}{\langle I \rangle}$. 

$E_c$ reduces to $k_{\rm B}T$ when the fluctuation theorem applies. After straightforward algebra,
the linearity $f(\epsilon)$ with $\epsilon$ is obvious for Gaussian fluctuation. Using the relation (\ref{sigmaItau}) $\sigma(I_\tau)^2=\sigma(D_\tau)^2=\frac{t_D}{\tau}\sigma(D)^2$, the definition of $E_c$ can be rewritten:
\begin{equation}
E_c\dot=\frac{\tau}{2}\frac{\sigma(I_\tau)^2}{\langle I \rangle}=\frac{t_D}{2}\frac{\sigma(D)^2}{\langle D\rangle}.
\label{EcDef}
\end{equation}
Note that $E_c$ as well as $f(\epsilon)$ are independent of $\tau$ and become characteristic of the dissipation. In many systems where the dissipation mechanisms occur uniformly at small scales in the bulk. $D$ is a sum over a large number of the small scale dissipative structures contained in the entire volume. Then $\langle D\rangle$ is proportional to the volume of the system and one expects also $\sigma(D)^2$ proportional to the volume of the system as well by virtue of the law of large numbers. Hence $E_c$ is intensive. It is more precisely a {\it characteristic energy density}. In the following, we refer to the relation (\ref{EcDef}) as the {\it moments convergence formula}. We stress that none of the previous arguments refer to a thermostat. The definition of $E_c$ can be apply to {\it athermal} systems as well.

\subsection{Remark on the central limit Theorem applied to $I_\tau$}
\label{ThCL}

One considers $\tau\longrightarrow \infty$. 
The variable of interest can be discretized as $I_\tau=\frac{1}{N}\sum_{i=1}^N I_i$, where $X_i=\int_{t'_i}^{t'_i+t_X}X(t)dt$ with a time step $t_X$. Hence, the smoothed variable $I_\tau$ results of a sum over a large number $N=\frac{\tau}{t_I}$ of independent random variables $I_i$, of identical average $\langle I \rangle$ and identical standard deviation $\sigma$. The Central Limit Theorem (CLT) states that, for $N$ large enough, $I_\tau$ are Gaussian whatever the statistics of $I_i$, with an average $\langle I \rangle$ and a standard deviation $\sigma (I_\tau)=\frac{\sigma}{\sqrt{N}}$. In that spirit, the equation (\ref{EcDef}) can be also viewed as a consequence of the CLT as, indeed, $\sigma (I_\tau)^2=\frac{\sigma^2}{N}$ with $N=\frac{\tau}{t_I}$. Hence the asymmetric function $f(\epsilon)$ reduces to:

\begin{equation}
f(\epsilon)=\frac{2}{t_I }\frac{\epsilon \langle I \rangle}{\sigma^2}
\label{AssFunIII}
\end{equation}
and
\begin{equation}
E_c=\frac{t_I}{2}\frac{\sigma^2}{\langle I\rangle}.
\label{EcTCL} 
\end{equation}
By definition of the correlation time, one can expect that the fluctuations of $I_i$ mimic the fluctuations of $I$ i.e. $\sigma\propto \sigma(I) $ and by the use of the stationary properties:
\begin{eqnarray}
E_c&=\frac{t_I}{2}\frac{\sigma(I) ^2}{\langle I \rangle}\\
&= \frac{t_D \sigma(D) ^2}{2\langle D \rangle}. 
\label{AssFunDII}
\end{eqnarray}
Rigorously, the CLT holds only for fluctuations of order of $\sigma(I_\tau)$ and not for large deviations \cite{Feller}. Nevertheless, one can first consider that $I_\tau$ is Gaussian as an experimental fact. Indeed in many experiments Gaussian shape are reported for the smoothed variables \cite{Ciliberto1,Bustamante2,Eckhardt04}. Moreover the experimental tests of the Fluctuation Relation requires a significant number of negative events of the smoothed injected power to check the probability ratio in (\ref{AssFundef}). Most of the time such tests implies systems with large fluctuations and small average value of the injected power. Hence the CLT should apply for a significant range of smoothing time $\tau\gg t_I$.\\ 

Finally, one should underline that the fluctuation theorem address fluctuations around $0$ of the smoothed injected power, which are usually extremely rare. In contrast, $E_c$ characterizes typical fluctuations on $I_\tau$. Indeed in the definition (\ref{EcDef}) the value $0$ does not play any specific role. Moreover, for a Gaussian process, the probability ratio introduced in ({\ref{AssFundef}), can be deduced for any $I_o\neq\langle I \rangle$. In such case the new asymmetric function becomes:

\begin{eqnarray}
f_{I_o}(\epsilon)\doteq &\frac{1}{\tau}\log\frac{P(I_{\tau}=I_o+\epsilon)}{P(I_{\tau}=I_o-\epsilon)}\\
&=\frac{2}{\tau}\frac{( \langle I \rangle-I_o) \epsilon}{\sigma (I_\tau)^2}\\
&=f(\epsilon)-\frac{2}{\tau } \frac{I_o \epsilon}{\sigma (I_\tau)^2}
\end{eqnarray}
As (\ref{sigmaItau}) still holds, one has $\frac{1}{\tau}\frac{I_o \epsilon}{ \sigma (I_\tau)^2} = \frac{1}{t_{\rm D}}\frac{ I_o \epsilon}{ \sigma(D) ^2}$ 
and $ \frac{1}{t_{\rm I}}\frac{\epsilon \langle I \rangle}{\sigma(I) ^2}=\frac{1}{t_{\rm D}}\frac{\epsilon \langle D \rangle}{ \sigma(D) ^2}$ 
independently. Actually the choice of $I_o = 0$ does not seems to play any specific role here. Moreover, the fact that any $|I_o| <\sigma(I_\tau)$ can be used ensuring the applicability of the CLT. In return that gives a certain consistency to the Gaussian fluctuations hypothesis and to the use of the relation of the moments convergence (\ref{EcDef}) to characterize the typical fluctuations of the injected and dissipated powers. \\

To summarize this section, we proposed to study of the fluctuations of injected power around the average, which are the easiest to access experimentally. These fluctuations give also relevant information about the intrinsic dissipative processes involved in the system, and allows to determine a characteristic density of energy. Moreover this approach does not need the concept of thermostat: it can be applied to an out-of-equilibrium system in contact with an energy reservoir as well as an athermal intrinsically dissipative system. In the former case, if the fluctuations are Gaussian, it has to be equivalent to its Fluctuation Relation counterpart. In the next section, we will illustrate these fact on Langevin equations describing either a particles in a thermal bath pushed out-of-equilibrium or an intrinsically dissipative system driven by random forcing.

\section{Application to damped particles driven by a random force} 
\label{Langevin}

\subsection{Case 1: A forced system in contact with a thermostat}
\label{C1}

In the usual framework of the Fluctuation Relation an the Fluctuation Theorem, a Brownian particle is pulled in a thermal bath, by an optical trap for instance \cite{SergioArtem}. The motion of the particle is given by: 

\begin{equation}
m\frac{dV}{dt}=-\gamma V +F(t) +f(t)
\label{BrII}
\end{equation}
where $m$ is the mass of the particle, $F(t)$ is the force exerted by the external operator whereas $f(t)$ is induced by the random action of the thermostat, i.e. it is usually a random white noise. When $F(t)=0$ the Einstein relation between $\gamma$ and $f$ holds :

\begin{equation}
\gamma=\frac{K}{m\langle \Delta V^2 \rangle}=\frac{\sigma(f)^2 t_f}{m\langle \Delta V^2 \rangle}
\label{EinstRel}
\end{equation}
with the diffusion coefficient $K=\int_0^{+\infty} \langle f(t+\tau) f(t)\rangle d\tau=\sigma_f^2 t_f $ 
and $\Delta V$ is the velocity of the particle without forcing. By virtue of the equipartition of energy at equilibrium, one can deduce that $\langle\Delta V^2 \rangle = \frac{d k T}{ m}$ where $d$ is the space dimension 
in which the Brownian particle evolves. 
In the presence of an external forcing, an additional velocity occurs: $V=V_F+\Delta V$. It may have a kind of ambiguity in the choice of the injected and dissipated power. Following Kurchan, if the injected power remains: 
$I=F(t) V$. Then to satisfy the balance (\ref{EnBal}), one has to choose the dissipation $D=\gamma V^2-f(t)\cdot V(t)$\cite{Kurchan}. 
Before computing the first moment of $D$, one notes that $F$ or $V_F$ can vary with time but are prescribed by the operator. They have to be the same in all samples used to perform the ensemble average. Therefore the averaging procedure must include an ensemble average $\langle \cdot \rangle$ concerning the thermalized variables, and time average $\overline{~\cdot~} $ for variables managed by 
the operator. One has: $\overline{\langle D \rangle}=\gamma\overline{V_F^2}$. 
To proceed, consider that (\ref{BrII}) 
characterizes an Ornstein-Uhlenbeck process, where 
$f(t)$ is itself described by a Langevin equation:
\begin{equation}
\frac{d f}{dt}=-\eta f(t) +\xi(t)
\label{OU}
\end{equation}
with $1/\eta$ the characteristic decay time of the random force $f$ and $\langle \xi(t)\xi(t+\tau) \rangle=2\Gamma \delta(\tau)$. In such case, the statistics of $f$ and $\Delta V$ follow a bivariate normal distribution with a Probability Density Function (PDF) \cite{Feller}:
\begin{equation}
P(\Delta V,f)=\frac{1}{2\pi\sqrt{1-r^2}\;\sigma(\Delta V)\cdot\sigma(f)}\exp\left[\frac{-1}{2(1-r^2)}\left(\frac{\Delta V^2}{\sigma(\Delta V)^2}-\frac{2 r \Delta V\cdot f }{\sigma(\Delta V)\cdot\sigma(f)}+\frac{f^2}{\sigma(f)^2}\right)\right],
\label{Bivariate}
\end{equation}
where 
$r=\frac{\langle f\;\Delta V\rangle}{\sigma(\Delta V)\cdot\sigma(f)}$ is called the correlator. From equations (\ref{BrII}, \ref{OU}, \ref{Bivariate}), we 
deduce:
\begin{equation}
\sigma(f)^2=\frac{\Gamma}{\eta};
\label{f2}
\end{equation}
\begin{equation}
\langle\Delta V\; f\rangle=\frac{\Gamma}{\eta(m\eta+\gamma)};
\label{fV}
\end{equation}
\begin{equation}
\sigma(\Delta V)^2=\frac{\Gamma}{\gamma\eta(m\eta+\gamma)}.
\label{V2}
\end{equation}
Moreover due to the Gaussian
character of the symmetric distribution (\ref{Bivariate}), we know that 
$\langle \Delta V^p \; f^q\rangle=0$ whenever $p+q$ is odd, and that
\begin{equation}
\langle\Delta V^3\; f\rangle=3\sigma(\Delta V)^2\langle\Delta V\; f\rangle
\label{fV3}
\end{equation}
\begin{equation}
\langle\Delta V^2\; f^2\rangle=\sigma(\Delta V)^2\;\sigma(f)^2+2\langle\Delta V\; f\rangle^2
\label{f2V2}
\end{equation}
Using these Gaussian properties
\cite{Feller}, and properties of the Fourier transform, we can deduce the Power Spectral Density (PSD) at vanishing frequency:
\begin{eqnarray}
\label{PSDD}
|\widehat{D}(\omega=0)|^2&=&\int_0^{\infty}\overline{V_F(t)\cdot V_F(t+\tau)}\cdot\langle f(t)\cdot f(t+\tau)\rangle d\tau+\\ \nonumber
&&\int_0^{\infty}\langle f(t)\cdot f(t+\tau)\rangle\langle \Delta V(t)\cdot \Delta V(t+\tau)\rangle d\tau-\gamma^2\int_0^{\infty}\langle \Delta V(t)\cdot \Delta V(t+\tau)\rangle^2 d\tau
\end{eqnarray}
where $\hat{X}(\omega)$ denotes the Fourier transform of $X(t)$. The right hand side of equation (\ref{PSDD}) can be integrated using the Fourier transform of (\ref{BrII}) and (\ref{OU}) to get the PSD of $f(t)$ and $\Delta V(t)$:
\begin{eqnarray}
\label{PSDfV1}
|\hat{f}(\omega)|^2&=&\frac{2\Gamma}{\eta^2+\omega^2}\\
|\Delta\hat{V}(\omega)|^2&=&\frac{2\Gamma}{(\eta^2+\omega^2)(\gamma^2+m^2\omega^2)}.
\end{eqnarray}
Using the Wiener-Khinchine theorem, replacing $\Gamma $ by $\sigma(\Delta V)^2\gamma\eta(m\eta+\gamma)$ and reminding that $\int_0 ^{\infty}e^{i\omega t}dt=\frac{1}{2} \delta(\omega)$
we can show that the two last terms of the right hand side of (\ref{PSDD}) cancel out. Dividing by $\overline{\langle D\rangle}=\gamma \overline{V_F^2}$, one gets the characteristic energy density of an out-of-equilibrium Brownian particle submitted to thermal force with exponential decorrelation.
\begin{equation}
E_c= \frac{|\hat{D}(0)|^2}{\langle D\rangle}=2\sigma(\Delta V)^2\frac{(m\eta+\gamma)}{\eta \overline{ V_F^2} }\int_{0} ^{\infty}\int_{-\infty} ^{\infty}\frac{\overline{V_F(t)\cdot V_F(t+\tau)}\cdot e^{i\omega t}}{1+\omega^2/\eta^2} d\omega d\tau
\label{Ec0}
\end{equation}
In the white noise limit where $\eta\rightarrow\infty$ (keeping $\Gamma/\eta^2$ constant), the usual result 
$E_c=m\sigma(\Delta V)^2$ is recovered \cite{Kurchan}. Note that equation (\ref{Ec0}) gives the correction of the Fluctuation Relation in the case of a correlated thermal noise. It shows that in this case, the driving velocity has to be taken into account in the characteristic energy density. 

\subsection{Case 2: A randomly driven athermal system}
\label{C2}

We apply the previous equations to the most elementary out-of-equilibrium intrinsically dissipative system described by:
\begin{equation}
m\frac{dV}{dt}=-\gamma V + F(t)
\label{BrI}
\end{equation}
where $F(t)$ is an external forcing, and $V$ is 
the velocity of a macroscopic particles moving in a viscous fluid, or the velocity of a thin elastic plate forced by a large shaker, 
a system to be discussed below (see \ref{ExpS1}).
Consider now $F(t)$ is a random forcing that does not include thermal fluctuations that are irrelevant for such a macroscopic system. (\ref{BrI}) does not describe anymore the motion of a Brownian particle in contact with a thermostat pulled out of its equilibrium state by an external applied force. Indeed, here as soon as the random forcing ceases the system goes at rest. Here we have: $I=V F$ 
and $D=\gamma V^2$. Assuming a Gaussian forcing, one expects a Gaussian distribution of velocity in the stationary regime. Therefore $\langle D \rangle =\gamma \langle V^2 \rangle$ and $\sigma(D)^2=\gamma^2(\langle V^4\rangle-\langle V^2\rangle^2)=2 \gamma^2 \langle V ^2\rangle^2$. Moreover one can compute that $t_D$, defined as $|\widehat{D}(\omega=0)|/\sigma(D)^2$ 
is exactly $m/\gamma$ . All that together with equation (\ref{AssFunDII}), it gives:
\begin{equation}
E_c= m\langle V^2 \rangle
\label{ThFCL}
\end{equation}
\
All the preceding reasoning does not precise the internal energy, that can include any potential energy term. We are concerned only by the budget of energy injection and dissipation. Hence the extension of the preceding results to systems with conservative force is straightforward. Moreover, the result $E_c\propto \langle V^2\rangle$ is strongly related to the form of the damping term: $\gamma V$ in equation (\ref{BrI}) which is suitable for any out-of-equilibrium system. Nevertheless, it is reasonable to apply this equation to describe the energy 
supplied to an electromagnetic shaker generating waves in a thin elastic plate. 
This study is presented in the following section.

\section{Experimental test}
\label{Exp}
\subsection{Experimental device}
\label{ExpDevice}
The experimental setup, sketched in figure \ref{ExpDev}, is similar to that commonly used to study wave turbulence in thin elastic plates\cite{Mordant}. It is made up a stainless steel plate of $2{\rm m}\times1{\rm m}\times0.5{\rm mm}$, rigidly fixed at the top. 
Waves are generated by a large electromagnetic shaker S1 (LDS V406). A generator (Agilent 33220A) produces a random noise windowed between $f_1=70\,$Hz and $f_2=90\,$Hz by a band pass filter (SR 650), finally amplified through a (Br\"uel \& Kj\ae r 2719) power amplifier before supplying the shaker S1. 

Following a general scheme of nonlinear `cascade', the input energy injected in this range is transferred up to higher frequencies (smaller wavelengths) where all the energy 
is dissipated. 

Different mechanisms of dissipation are involved along the cascade depending on the wavelength \cite{Humbert}. This is typically an `athermal' out-of-equilibrium dissipative system, for energy must be supplied to sustain
a steady state motion of the plate. 
The nonlinearly coupled modes generated in the plate reach a turbulent state extending between the forcing scale ($\lambda_o\sim 250\,$mm) to the smallest scales ($\sim h=0.5\,$mm). 
A large number of degrees of freedom results from this `turbulent state' $N\sim(\lambda_o/h)^2=2.5\,10^5$. In addition to this large shaker driving the plate motion, we use a smaller one, S2, located at a distinct position (see figure \ref{ExpDev}). This small shaker (Br\"uel \& Kj\ae r 4810) is attached to the plate, separately excited by S1. This small shaker is driven by an Arbitrary Wave Generator (Agilent 33522A) through power Amplifier (NF electronic instruments 4005 High speed amplifier). Hence, it also supplies energy to the plate but at a much smaller rate (from 50 to 1000 times smaller). It is typically excited by a small sinusoidal current at a frequency $f=33~$Hz. 

The force applied by the large shaker S1 is measured with strain Gauge (Testwell~KD40S). The velocity of the plate at the shaker position is estimated with a Laser Vibrometer (Polytec~OFV505). Hence, we measure the power injected by S1 by multiplying the time series of force and velocity acquired by a 24 bits A/D converter (NI PXI 4462). 

We use an electrical ansatz to measure the power injected by S2. 
Remembering that the shaker is nothing but a coil moving in a permanent magnetic field, we calculate the electromotive force (emf) $e$ induced by the motion. This motion is caused on one hand by the current $i$ sent in she shaker, and the plate's vibrations on the other hand. The emf reflects the shaker's coil velocity, i.e. the velocity of the plate at the point of contact, whereas the current mimics the force applied on the plate by the shaker. Following the sketch figure \ref{ExpDev}-B, one has $i=(Uo-U_1)/R$ and $e= U_1-r i$. 

A symmetry of electromagnetic laws implies that the prefactor between $e$ and $V$ on one hand and that between $i$ and $F$ on the other hand are inversely related, such that $Fv=i\,e$. } Therefore, $I_e=Fv=i\,e$ pictures the power injected by S2 into the plate. (The same procedure has been used to probe granular gases\cite{Naert}, however with a DC motor. The internal resistance measured with an ohmmeter $ r \ simeq2 \ pm0.2 ~ \ Omega $ is sensitive to temperature variation and to the piston pi. In order to reduce the signal-noise we artificially increase it such that $ r \ simeq26.4 \ pm0.2 ~ \ Omega $. Besides, we checked that the inductance is negligible, which is natural at such a low frequency. 

\begin{figure}[h!]
\centering
\resizebox{0.9\textwidth}{!}{%
\includegraphics[width=7in]{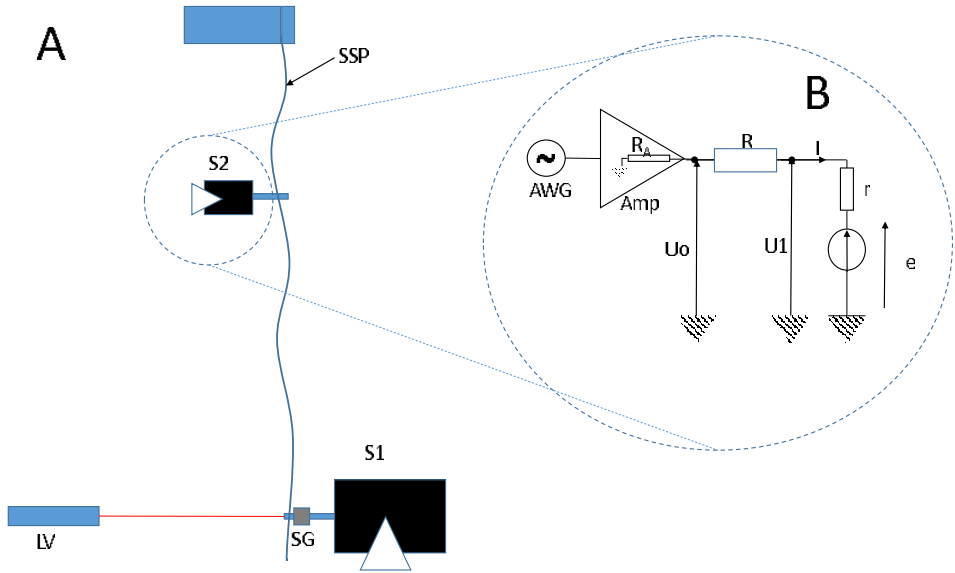}
}
\caption{The experimental device. {\bf A}: side view of the thin Stainless Steel Plate (SSP) 
excited by an electromagnetic shaker S1. The Strain Gauge (SG) measures the applied force and the Laser Vibrometer (LV) measures the velocity at the Shaker position. A small amount of energy is also supplied by the shaker S2. \\
{\bf B}: Design of the electrical 
circuit feeding S2. The current is provided by an Arbitrary Wave Generator (AWG) and amplified (Amp). $U_o$ and $U_1$ are measured around the shunt resistor $R$. It gives access to the current $i$ and the emf $e$ knowing the internal resistor $r$ of the shaker S2.}
\label{ExpDev}
\end{figure}
\noindent The motion of the plate at the point of contact of S1 is described in first approximation by an equation like (\ref{BrI}), where $F(t)$ is a random forcing. 
One must include on the RHS, however, a restoring elastic force induced 
by a spring in the shaker, which serves to guide the moving coil in translation. Although it could be non-linear for displacements wider than a few mm,
this restoring force has no impact on the power budget as long as it remains conservative because it leaves $I$ and $D$ unchanged. \\

We expect therefore $Ec_1\propto \langle V_1^2\rangle$ with $V_1$ the velocity of the plate measured at the at the point of contact of S1. However, even if (\ref{BrII}) can be applied to describe the dynamics of S2, the relation (\ref{EinstRel}) does not hold 
because the exchange with S2 is intrinsically dissipative.

We must underline that setting the AWG voltage to $0$ does not ensure that the current imposed to S2 is actually $0$. Indeed, each time the shaker S2 generates a voltage, because it is set in motion by the plate, a current flows and dissipates the energy in the all the resistors ($r$, $R$ and $R_A$). This generates a negative offset to the injected power that one must remove. Indeed the mean electrical input power can be negative for a small forcing if it does not overcome this intrinsic dissipation. Hence for each excitation of the large shaker, we measure this offset by setting the AWG to 0. Then it is subtracted from the average injected power measured when a current is supplied by the AWG. This pumping action of the small shaker underlines its intrinsically dissipative nature. \\
The figure \ref{PDFI1} shows that for small $\tau$, the 
fluctuations of $I_{1\tau}$ are far from 
Gaussian.
The histograms keep reminiscence of the shape of the unsmoothed power. This PDF can be computed exactly for a force and a velocity that are two correlated Gaussian variables of zero mean, like in the situation discussed here\cite{FalconWT}. 

\begin{figure}[!h]

\centering
\resizebox{0.9\textwidth}{!}{%
\includegraphics[width=7in]{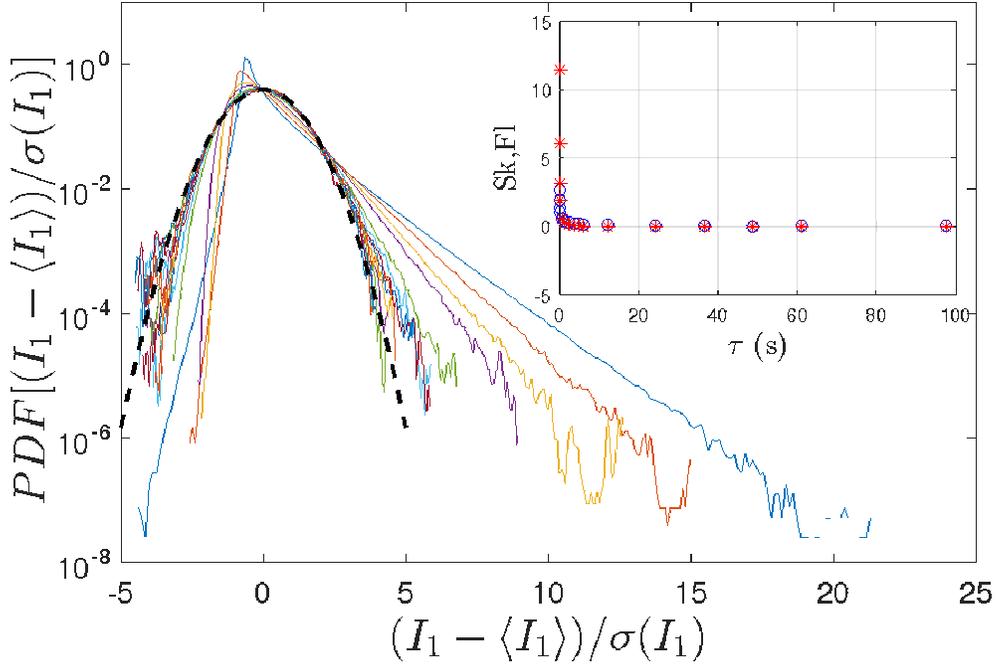}
}

\caption{Main panel: Probability Density Function of the smoothed power, $I_{1\tau}$, injected by the large shaker S1,. Inset: The skewness $Sk=\langle (I_{1\tau} -\langle I_{1\tau}\rangle)^3\rangle/\sigma( I_{1\tau})^3$ and the flatness : $Fl=\langle (I_{1\tau} -\langle I_{1\tau}\rangle)^4\rangle/\sigma( I_{1\tau})^4$ minus 3 of the smooth injected power, as a function of the smoothing time $\tau$. }
\label{PDFI1}
\end{figure}

For $\xi(t)$ in (\ref{BrI}) being a Gaussian white noise, the 
PDF of the smoothed variable $I_{1\tau}$ can be also computed exactly \cite{Farago}. Nevertheless, by virtue of the central limit theorem, the PDFs become Gaussian as testified by the skewness and the flatness that converge to $0$ and $3$ respectively, when $\tau$ overcomes $10~$ms. Concerning $I_{2\tau}$, fluctuations are always close to the Gaussian as shown figure \ref{PDFI2}. The skewness is nearly $0$ and the flatness never exceeds $4.5$ (compared to $13$ for $I_{1\tau}$). Both converge quickly to the Gaussian values.\\
We note here, from the experimental data, that the fluctuations of injected and dissipated powers have very different statistics.

\begin{figure}[!h]

\centering
\resizebox{0.9\textwidth}{!}{%
\includegraphics[width=7in]{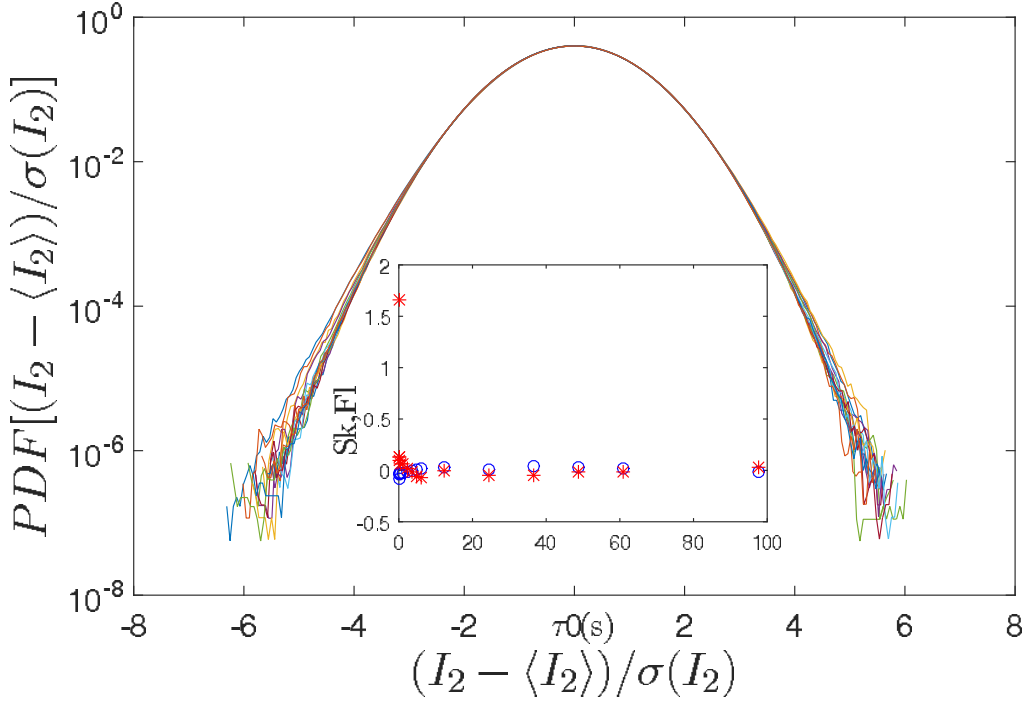}
}

\caption{Main panel: Probability Density Function of the smoothed power, $I_{2\tau}$, injected by the small shaker S2. Inset: The skewness 
$Sk=\langle (I_{2\tau} -\langle I_{2\tau}\rangle)^3\rangle/\sigma( I_{2\tau})^3$ and the flatness minus 3 : $Fl=\langle (I_{2\tau} -\langle I_{2\tau}\rangle)^4\rangle/\sigma( I_{2\tau})^4-3$ of the smooth injected power as a function of the smoothing time $\tau$.
}
\label{PDFI2}
\end{figure}

\subsection{Experimental results}
\label{Result}
\subsubsection{The large shaker S1}
\label{ExpS1}

\begin{figure}[h!]

\centering
\resizebox{0.9\textwidth}{!}{%
\includegraphics[width=7in]{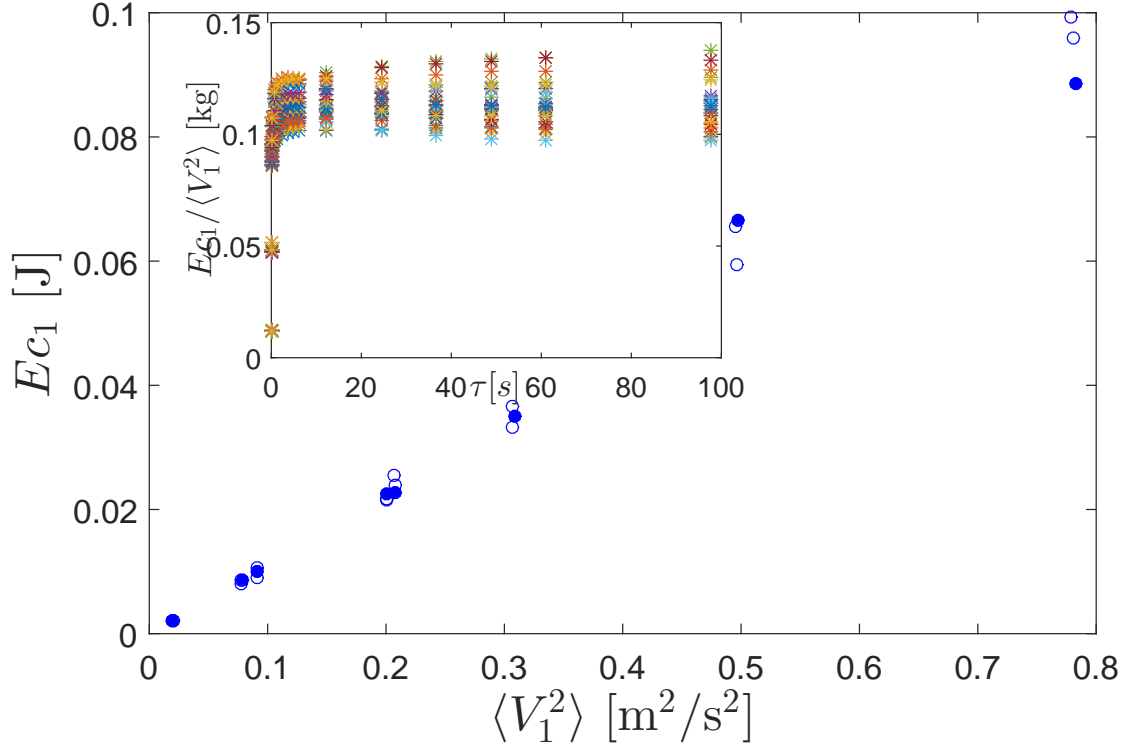}
}

\caption{{\bf Main panel}: the coefficient $Ec_1\dot=\tau\sigma(I_{1\tau})^2/(2\langle I_1\rangle)$ as a function of the variance of the shaker velocity. $Ec_1$ is estimated for the five larger $\tau=[24.4 ; 36.6 ; 48.8 ; 61.0 ; 97.7]$s (shown in the inset). Blue Bullet no current is injected in S2, open circle some power is injected through S2.{ \bf Inset}: The ratio $Ec_1/\langle V_1^2\rangle$ as a function of the smoothing time $\tau$ shows a fast converging to a constant value above $\tau=10$ s. }
\label{FlucRelS1}
\end{figure}

We check on figure \ref{FlucRelS1} that $Ec_1$ defined as $Ec_1\dot=\tau\sigma(I_{1\tau})^2/(2\langle I_1\rangle)$ is indeed proportional to $\langle V_1^2\rangle$ at large $\tau$, as expected from Equation \ref{EcTCL}. The data are obtained with different forcing configuration of both shakers S1 and S2. Note that S2 affects poorly the behaviors of S1. $\tau$ ranges from $0$ to $100\,$s. The inset shows that after $20\,$s, $Ec_1/\langle V_1^2\rangle$ reaches a constant. This is $2000$ correlation times $t_{I_1}$ extracted from the low frequency limit of the Power Density Spectrum (PSD) which gives $t_{I_1}\sim10\pm0.5$ ms, whatever is the forcing. The constant reached by $Ec_1/\langle V_1^2\rangle$ is assumed to be the effective mass moved by the shaker. The small spreading may be explained by the small amount of energy exchanged with the small shaker S2.

\subsubsection{The small shaker S2}
\label{ExpS2}
\begin{figure}[h!]
\centering
\resizebox{0.9\textwidth}{!}{%
\includegraphics[width=7in]{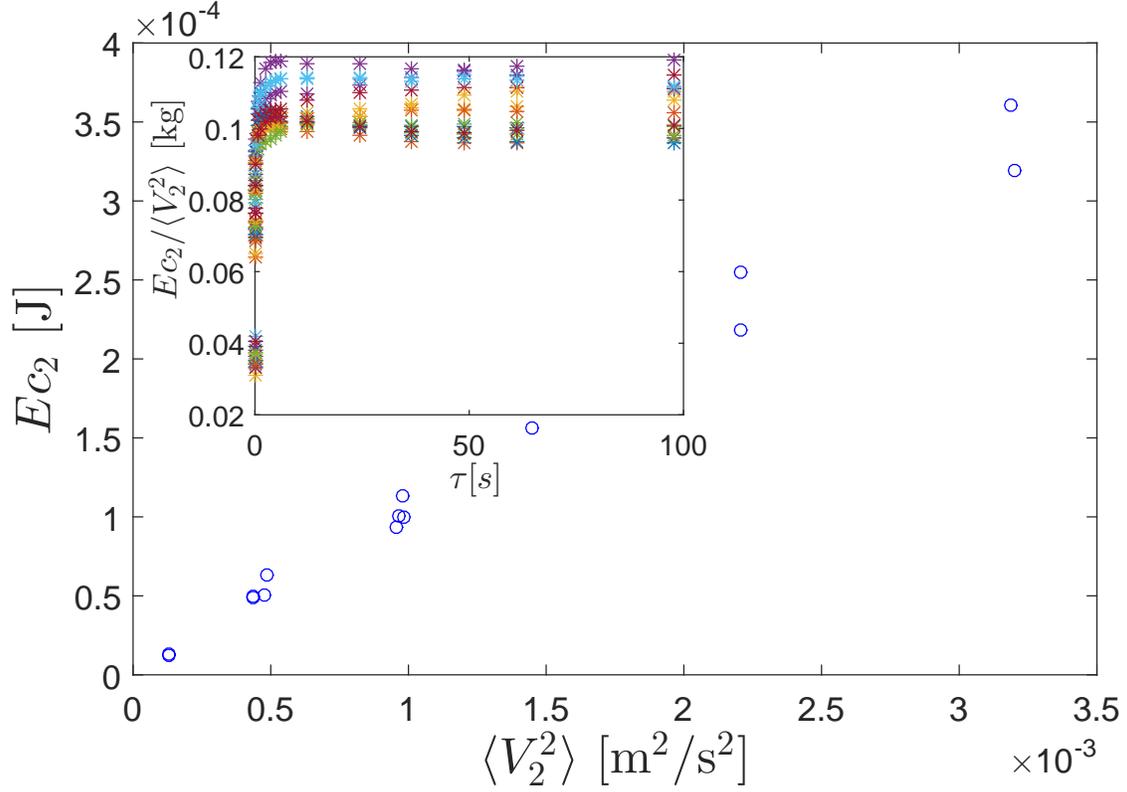}
}
\caption{{\bf Main panel}: the coefficient $Ec_2\dot=\tau\sigma(I_{e\tau})/[2(\langle I_e\rangle-\langle I_o\rangle)]$ as a function of the $\langle \Delta V^2\rangle= \langle V_2^2\rangle+\langle V_o^2\rangle$. $Ec_2$ is estimated for the five larger $\tau=[24.4 ; 36.6 ; 48.8 ; 61.0 ; 97.7]$s (shown in the inset). {\bf Inset}: The ratio $Ec_2/\langle\Delta V_2^2\rangle$ as a function of the smoothing time $\tau$ shows a fast converging to a constant above $\tau=10$ s.}
\label{FlucRelS2}
\end{figure}

Now let's check whether, for the shaker S2, the following relation holds:
\begin{equation}
Ec_2\dot=\frac{\tau}{2}\frac{\sigma(I_{2\tau})^2}{\langle I_2\rangle-\langle I_o\rangle} \propto\langle V_2^2\rangle
\end{equation}
with $I_2$ the power injected into the shaker S2 whereas $I_o$ is the offset power pumped from the plate motion by the device without external voltage supplied into S2 and $V_2$ being the velocity of the piston. Figure \ref{FlucRelS2} shows also a good proportionality between $Ec_2$ and $\langle V_2^2\rangle$. It convergences to an asymptotic value for a smoothing time $\tau$ around $10\,$s or smaller. This value corresponds also to an effective moving mass. It is of the same order than the one extracted for S1. The meaning of these masses deserves further investigations. Note that the correlation time of $I_e$, is about $3~$ms, hence $3\times10^3$ smaller than the value of $\tau$ where the relation of the moments convergence is fulfilled. 
We presented in this section the experimental study of the power injected in vibrating elastic plate, excited in a nonlinear regime by two shakers. The two shakers play different roles: the large one maintains the motion of the plate that would otherwise remain at rest, while the smaller one, used as a probe, is subjected to an external forcing in addition to the motion of the plate. Our approach consists in analyzing the fluctuations of power exchanged through these shakers between the plate and the external world. We apply the same analytical framework at each shaker. It reveals two distinct characteristic energies as defined by (\ref{EcDef}). In both case, these energies are proportional to the velocity variance at the contact-point between the plate and the corresponding shakers. Actually, this illustrates that both shakers are subjected to a viscous damping either induced by the waves generated in the plate or due the internal dissipation of the shakers and the measurement tools. In the following section we check if the linear relation between characteristic energy and velocity variance survives in the case of frictional systems.

\section{The relation of the moments convergence for model of nonlinear sliding blocks}
\label{BK}
In the example discussed above, the viscous damping allows us to assimilate the relation of the moments convergence to a fluctuation-like relation. 
To show that this protocols is indeed quite general, we investigate the relation of the moments convergence for a system with a different kind of dissipation. The non-conservative friction force appears in many other mechanical systems. The discontinuity and the nonlinearity of the sliding friction are known to enrich the dynamics of Brownian motion \cite{TouchetteI,TouchetteII}. Because of the limited duration of translational friction experiments and the periodicity induced in rotational frictional setup \cite{Baumberger94,Geminard2016}, testing our long-time prediction experimentally is difficult. 
We then turned to numerical simulations of the Burridge-Knopoff model of nonlinear sliding blocks \cite{BK}. In this model, identical blocks attached to each other by identical springs are pull ahead at constant velocity $V_0$, and the last block follows the previous one freely, as depicted on figure \ref{BKdyn}-left. 

\begin{figure}[!h]
\centering 
\begin{minipage}[c]{.95\linewidth}
\includegraphics[width=1\textwidth]{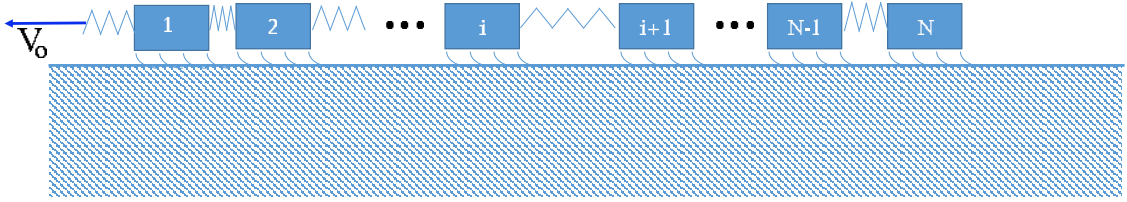}
\end{minipage}
\caption{Sketch of the Burridge-Knopoff model in use. The first spring is pulled at the constant speed $V_0$, whereas the last block follows freely the previous one.}
\label{BKdyn}
\end{figure}

The blocks slide only in the pulling direction with a nonlinear friction described by the last term of the following equations of motion of the i\textsuperscript{th} block:
\begin{eqnarray}
m\ddot{X_i}= k&\left(X_{i-1}-2\cdot X_i+X_{i+1}\right)-\frac{F_o}{1+a \dot{X_i}} \mbox{ if } \dot{X_i}>0\\
\dot{X_i}= 0 & \mbox{ elsewhere }
\label{BKeqMotion}
\end{eqnarray}
with $X_i$ the position of the block $i$, $m$ the mass of the blocks, $k$ the spring strength and where $F_o/(1+a\,{X_i})$ illustrates the sliding facilitation at high speed. It is parametrized by the characteristic velocity $1/a$. The other control parameter is the pulling speed $V_o$. Using the convenient time scale $\sqrt{m/k}$ and displacement scale $F_o/k$, the dimensionless equations becomes:
\begin{eqnarray}
\ddot{U_i}= & k\left(U_{i-1}-2\, U_i+X_{i+1}\right)-\frac{1}{1+\alpha \dot{X_i}} \mbox{ if } \dot{U_i}>0\\
\dot{U_i}= & 0 \mbox{ elsewhere }
\label{BKeqMotionDL}
\end{eqnarray}
where $U_i=X_i\,( k /F_o)$ and with the parameters
: $\alpha =a F_o/ \sqrt{m\, k}$ and the dimensionless pulling velocity $\mu_o=V_o\, \sqrt{k\, m}/F_o$.
We deliberately use a small number of blocks, $N=3$, because they already exhibit a very perturbed stick-slip motion, as shown in figure \ref{BKdyn}. It allows us to perform easily very long simulation in time. Moreover, we know that the addition of many blocks do not complexify to much the dynamical behavior \cite{Saumaitr}. Indeed, due to the self-similar behavior of the chains the main events of the dynamical equations (\ref{BKeqMotionDL}) correspond to avalanches involving all the blocks \cite{deSousaI,deSousaII}. The addition of blocks just slows down the dynamics and the statistical properties can be rescaled \cite{Saumaitr}.

\begin{figure}[!h]
\centering 
\includegraphics[width=1\textwidth]{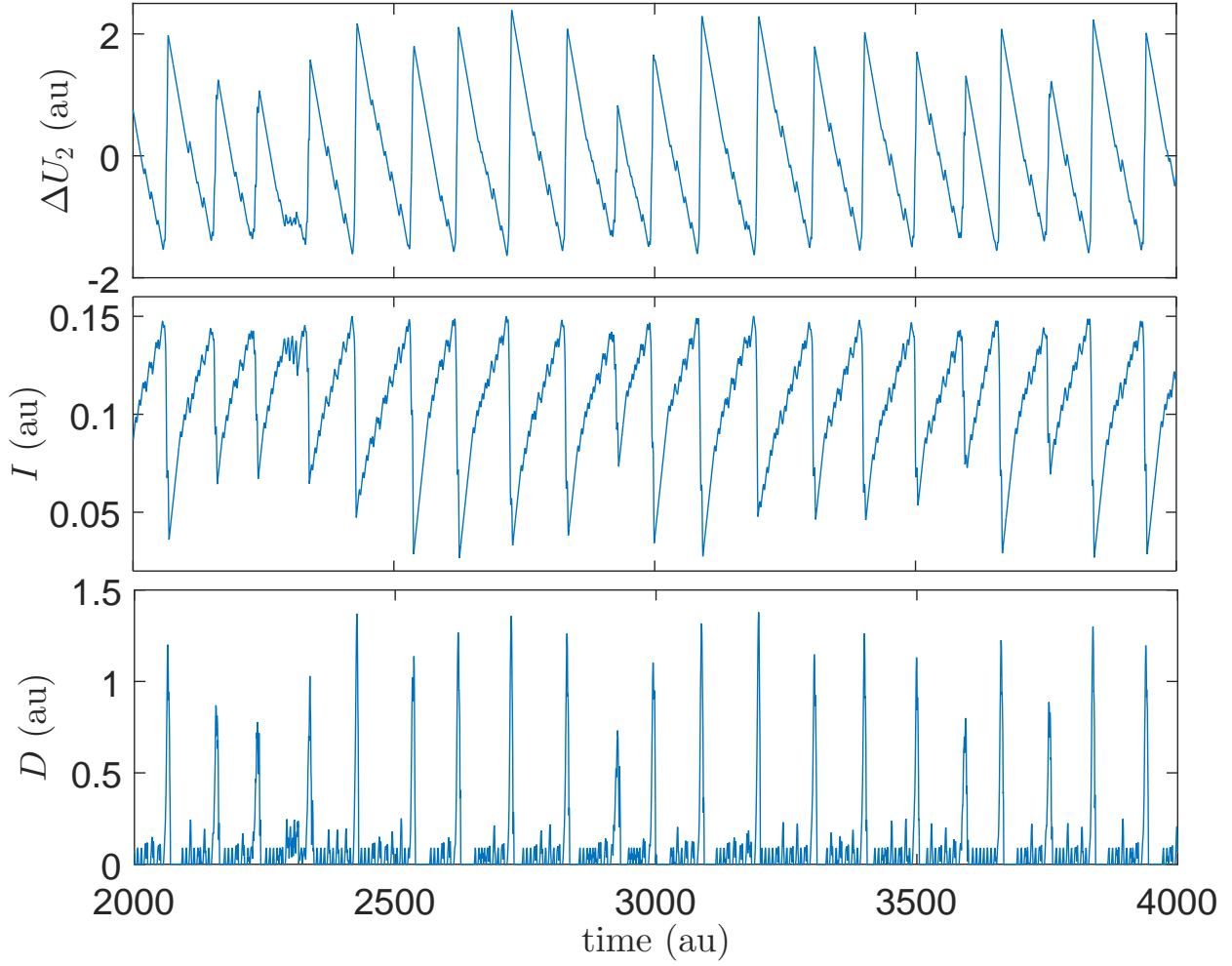}
\caption{
Temporal traces of the position of the second block in the co-moving frame (top),of the injected (middle) and dissipated powers (bottom) for $\alpha=1/2$ and $\mu=1/20$.}
\label{BKdyn}
\end{figure}
With only three blocks, the dynamics of global variables is fast enough to get a very good convergences of power density spectra of injected and dissipated power. Hence we can probe their statistical behaviors along smoothing time $\tau$ within a reasonable simulation time. The temporal traces of the power injected and dissipated differ a lot, as shown on the temporal traces on figure \ref{BKdyn}. The injected power $I = \mu_o (\mu_o\, t-U_1) $ reproduces mainly the opposite of the stick-slip motion of the first block in the co-moving frame. It fluctuates around its average with relative fluctuations of about 25\%. In contrast, the dissipated power is much more intermittent and is concentrated in many high peaks corresponding to a global sliding of the blocks. Its relative fluctuations are about 210\%. Nevertheless, although their spectra, represented in the figure \ref{PSDBK}, differ, their limits at low frequencies converge to the same value as expected for a stationary process. This proves that the characteristic energy density defined in (\ref{EcDef}) traces back to the dissipation also for this model.

\begin{figure}[!h]
\centering
\resizebox{0.9\textwidth}{!}{%
\includegraphics[width=5in]{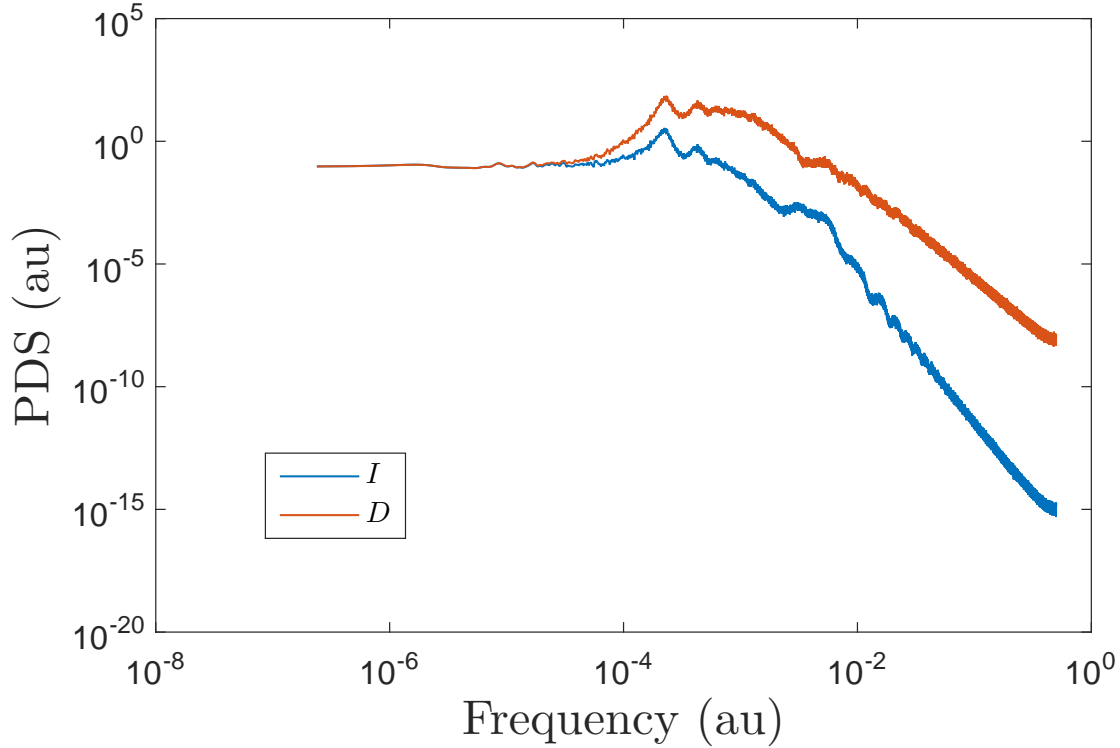}
}
\caption{Power Density Spectra of the dimensionless injected and dissipated power for $\alpha=1/2$ and $\mu=1/20$.}
\label{PSDBK}
\end{figure}
To go further, we test the convergence of the low frequencies limit of both PSD for various values of the two parameters of the model $\alpha$ and $\mu_o$. The main panel of the figure \ref{GK_BK} shows that the relation of the moments convergence holds very well for all tested parameters. Nevertheless the proportionality with the kinetic energy, expected for a viscous damping, is completely lost here, as shown in the lower inset of figure \ref{GK_BK}. The upper inset shows that the total energy cannot rescale the energy defined by equation (\ref{EcDef}). This clearly demonstrates that the dissipation mechanism drives the properties of the characteristic energy $E_o$. 
\begin{figure}[!h]

\centering
\resizebox{0.9\textwidth}{!}{%
\includegraphics[width=5in]{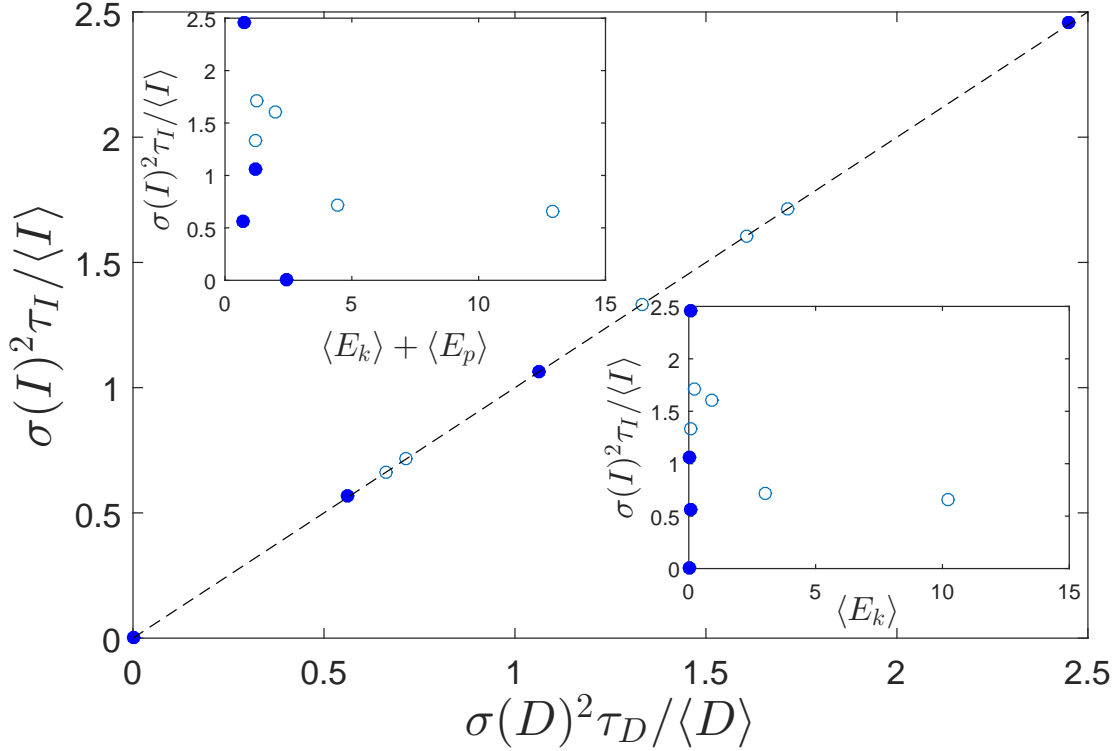}
}
\caption{\textbf{Main panel}: Test of the relation of the moments convergence (\ref{EcDef}) on the Burridge-Knopoff model. The blue open circles are obtained for $\alpha=1/2$ and $\mu_o=[2, 1, 1/2, 1/5, 1/10]$. The blue bullets are obtain for $\alpha=[1/10, 1/2, 1, 10]$ and $\mu_o=1/20$. \textbf{Upper inset}: 
$Ec=\sigma(I)^2\tau_I/\langle I\rangle$ as a function of the total energy for the same range of the parameters. \textbf{Lower inset}: 
$Ec=\sigma(I)^2\tau_I/\langle I\rangle$ as a function of the kinetic energy $E_k=\sum_i^3 U_i^2/2$, for the same range of the parameters. }
\label{GK_BK}
\end{figure}
We can conclude from these results that the characteristic energy density extracted from measurements of injected power is well suited to the fluctuations of the dissipative processes. Nevertheless, due to the complex form of the dissipation, it can not be related by any manner to the kinetic energy per block.

\section{Conclusions}
\label{Conclu}

We can summarize the canvas of this work as follow. First we recall that the fluctuation relation can be expressed with the two first moment of the injected power smoothed over a time $\tau$ for Gaussian fluctuations. We use this relation that defines a characteristic energy density as the starting point, because it is straightforward to relate it to dissipation as a consequence of the stationarity.
We call it the relation of the moments convergence (\ref{EcDef}). This relation links typical fluctuations of the extrinsic injection to the intrinsic dissipation. This approach presents several advantages. First it is not limited to systems in contact with a thermal bath. There is no need for the notion of temperature and it can be applied to power injected into athermal systems, whereas the Fluctuation Relation holds only for system in contact with a well define thermostat. 
The relation of the moments convergence makes use of the two first moments. One note that the presence of positive and negative fluctuations of the smoothed injected power is not required, as it is for the Fluctuation Relation. 
Often, collecting enough negative events is very difficult to catch experimentally at the large smoothing time limit, as prescribed by the Fluctuation Theorem. Moreover, we show that the relation of the moments convergence coincides with the Fluctuation Relation for a Brownian particle submitted to a thermal white noise, but it can be extended straightforward to correlated noise. Nevertheless the interpretations slightly differs.
Here the fluctuation relation is induced by the specific form of the viscous damping proportional to the velocity.
\\
To check the relevance of this relation of the moments convergence, we apply it to two different systems. The first one is an experiment where nonlinear waves are generated in a thin elastic plate by a large electromagnetic shaker. Another shaker, smaller, is attached to the excited plate. It is used as a probe and provides only a small quantity of energy compared to the one injected by the large shaker. Both shakers are intrinsically dissipative because they consume a part of the energy in internal impedance in addition to the energy provided to the plate. It turn out that this characteristic energy density extracted from the fluctuations of injected power for both shakers, is proportional to the variance theirs velocities in these cases. The proportionality factor which of the same order for both shaker, is related to the mass moved by the shakers. This is due to the fact that the overall dissipation can be well described by a viscous damping in this system. \\
The second system considered is the Burridge-Knopoff model. It is a numerical model of blocks attached to each other by identical springs and sliding forward with a nonlinear friction force when the first spring is pulled at constant speed. The relation of the moments convergence also holds in this case. However the characteristic energy density defined this way does not reduce to any kind of kinetic energy of the blocks because the dissipation cannot be reduced to a viscous damping.\\
The stationarity used here to relate injection and dissipation is a reminiscence of the first demonstrations of the fluctuation relation. Indeed an instantaneous stationarity was required in the sense that the energy was fixed constant by an artificial instantaneous equality of the injected and dissipated power \cite{EvansCohenMoriss}. However the interpretations differ. If the characteristic energy density defined here is proportional to a velocity variance, this is only a consequence of the viscous nature of the damping. The large deviation theory might be a promising way to connect in a more general way the relation of the moments convergence and the Fluctuation Theorem, the first being concerned by characteristic fluctuations of power around the mean and the second being usually concerned by rare events. 

Our approach is very general and applies to any dissipative systems in a stationary state. The specific case of turbulent flow is especially interesting from this point of view. Indeed the relation of the moments convergence seems to contradict strongly the simplest scaling model of turbulence due to Kolmogorov in 1941\cite{Frisch}. Within the K41 framework, using the definition $E_c= t_I\sigma(I)^2/(2\langle I \rangle)$ one expects $E_c\propto U^2$. In contrast, because the dissipation is a global quantity involving the smallest scales in the K41 framework, one expects $E_c\propto U^2/Re^{11/4}$ with the definition $E_c= t_D\sigma(D)^2/(2\langle D\rangle)$, $Re=U.L/\nu$ being the Reynolds number with $U$ a characteristic velocity of the flow, $L$ a characteristic length of the system and $\nu$ the fluid viscosity. On this basis, the hypothesis of this simplest scaling would necessitate a strong revision. It shows that the energy transfer from the large injection scale up to the dissipation deserves further investigations.

\end{document}